\documentclass{elsart}

\begin{document}
\begin{frontmatter}

\title{Nonlinear Dynamics of a Single Ferrofluid-Peak in an
       Oscillating Magnetic Field}

\author{Thomas Mahr}
\address{ Institut f\"ur Experimentelle Physik,
          Otto-von-Guericke-Universit\"at,
          Postfach 4120,
          D-39016 Magdeburg,
          Germany}

\author{Ingo Rehberg}
\address{ Institut f\"ur Experimentelle Physik,
          Otto-von-Guericke-Universit\"at,
          Postfach 4120,
          D-39016 Magdeburg,
          Germany}

\begin{abstract}

If a magnetic field normal to the surface of a magnetic fluid is
increased beyond a critical value a spontaneous deformation of
the surface arises (normal field instability). The instability
is subcritical and leads to peaks of a characteristic shape. We
investigate the neighborhood of this instability experimentally
under the influence of a temporal modulation of the magnetic
field. We use a small vessel, where only one peak arises.
The modulation can either be stabilizing or destabilizing,
depending on the frequency and amplitude.
We observe a cascade of odd-numbered response-periods
up to period 11, and also a domain of even-numbered periods.
We propose a minimal model involving a cutoff-condition which
captures the essence of the experimental observations.

PACS: 47.20.-k, 47.20.Ky, 75.50.Mm \\
Keywords: magnetic fluid; nonlinear oscillator; subharmonic response;
surface instability;

\end{abstract}
\end{frontmatter}

\section{Introduction}

Ferrofluids are colloidal suspensions of magnetic monodomains in a
non\--magnetic carrier liquid \cite{RosenBuch}. They behave like a
super-paramagnetic substance, which allows for a wide range of
applications \cite{Konferenz95}, and a peculiar flow behavior under
the influence of external magnetic fields
\cite{BacriSalin91,Sudo93,BacriElias96}. In general the hydrodynamics
of polarizable fluids is an interesting and non-trivial topic which
includes such subtle effects as counter propagation of free
surfaces under the influence of external magnetic fields
\cite{Rosensweig96,Liu}.
Experimental investigations of the flow dynamics are hindered
by the fact that the fluid is opaque. This seems to be the reason that
investigations of the dynamic behavior of magnetic fluids are sparse,
in spite of its tremendous technological potential \cite{applications}.
We feel that investigations of dynamic surface deformations, which can
be optically detected, are a practical approach to get a quantitative
measurement of the flow behavior.
In our experiment we drive nonlinear oscillations of the free surface
of a ferrofluid by making use of the technical advantage that
hydrodynamic motion can be induced by time dependent magnetic fields.
The magnetic driving is particularly efficient in the neighborhood of
static surface deforming instabilities, which can be achieved in
super-paramagnetic fluids at relatively small external fields.
These static instabilities are accompanied by hysteresis leading to a
particularly rich behavior of the nonlinear surface oscillations.
Here we present an experimental investigation of such field induced
surface deformations.
The resulting bifurcation scenario of the nonlinear surface
oscillations falls into a class which has never been studied
experimentally.
We discuss the results in terms of a qualitative
model which captures the main features of the experimental findings.

\section{Experimental Setup and Procedure}

The experimental setup is shown in Fig.~\ref{setup}.
In order to obtain a suitable compromise between viscosity and
magnetic permeability,
we use a mixture of the commercially available ferrofluids EMG 901
(Ferrofluidics) and EMG 909 in a ratio of 7 to 3.
The properties of EMG 901 are:
density $\rho = 1530$ kgm$^{-3}$,
surface tension $\sigma = 2.95 \cdot 10^{-2}$ kgs$^{-2}$,
initial magnetic permeability $\mu = 2.3$,
magnetic saturation $ M_S = 4.8 \cdot 10^4$ Am$^{-1}$,
dynamic viscosity $ \eta= 10 \cdot 10^{-3}$ Nsm$^{-2}$.
The properties of EMG 909 are:
density $\rho = 1020$ kgm$^{-3}$,
surface tension $\sigma = 2.65 \cdot 10^{-2}$ kgs$^{-2}$,
initial magnetic permeability $\mu = 1.8$,
magnetic saturation $ M_S = 1.6 \cdot 10^4$ Am$^{-1}$,
dynamic viscosity $ \eta= 6 \cdot 10^{-3}$ Nsm$^{-2}$.
By assuming a linear interpolation for the fluid parameters of the
mixture we obtain $\rho=1377$ kgm$^{-3}$, $\sigma=2.86 \cdot 10^{-2}$
kgs$^{-2}$, $\mu=2.15$, $M_S=3.84 \cdot 10^3$ Am$^{-1}$ and
$\eta=8.8 \cdot 10^{-3}$ Nsm$^{-2}$.
The critical field $H_c$ for the onset of the normal field instability
\cite{RosenBuch} has been measured to be $H_c = 6.2 \cdot 10^3$ Am$^{-1}$.
This is in reasonable agreement with the theoretical value
$H_c = 5.9 \cdot 10^3$ Am$^{-1}$ calculated for infinite fluid layer.
According to the same theory the critical wavelength is expected to
be 9 mm.
Due to aging of the fluid, presumably caused by evaporation of the
carrier liquid, the critical field changes within one week by about
5\%. The experimental runs shown below took on the order of hours,
where aging effects can be estimated to be on the order of 0.1\%.

The fluid is filled in a cylindrical teflon vessel of 12 mm depth and
3 mm diameter which is small in comparison to the critical wavelength
of the normal field instability
in a two dimensional system and thus enforces the existence of a
single peak. The upper 2 mm of the vessel has a slope
with respect to the horizontal
of $15^{\circ}$, which is close to the measured
contact angle between fluid and teflon, in order to provide a flat
surface of the fluid \cite{MaRe95}.

An image of the vessel is shown in Fig.~\ref{minmaxFoto}, where also
two snapshots of the ferrofluid peak are presented, the first one
taken at the phase of the oscillation where the amplitude reaches
its minimum value, the second one taken at the maximum amplitude.

The vessel is placed in the center of a pair of Helmholtz-coils
(Oswald), with an inner diameter of 40 cm. One coil consists of
474 windings of flat copper wire with a width of 4.5 mm and
thickness 2.5 mm. A current of about 5 A is then sufficient to
produce the magnetic field of about $8 \cdot 10^3$ Am$^{-1}$ used in this
experiment.
The static field is monitored by means of a hall probe
(Group 3 DTM-141 Digital Teslameter) located below the vessel.

The dynamics of the ferrofluid surface is detected by a digital line
scan camera (i2S iDC100) which is focused on the vertical axis
through the center of the vessel. The time between two exposures
is set to 6 ms.

The analysis of the lines is done with a 90 MHz Pentium-PC,
equipped with a 6 bit interface board (i2S ISM197) for the line
camera. The resolution in our experiment is 10.5 pixels per mm.
For controlling the experiment the PC is also equipped with a
synthesizer-board (WSB-10), and two programmable counters
(8253), located on a multifunction I/O-board (Meilhaus ME-30).
The counters are used to keep track of the the pacing-frequency of
the synthesizer-board. Their output is used to trigger the camera in
any desired phase of the driving oscillation of the magnetic field.
Thus we are using a phase-locked technique between the driving and
the sampling in order to ensure a jitter-free measurement of the
amplitude.
By keeping track of the synthesizer pace the computer moreover
manages the writing of the data into the synthesizer memory
at times where no conflict with the DA-converter arises. This allows
for smooth switching of the amplitude of the ac-component of the
magnetic field.
The wave-signal is amplified by a linear amplifier
(fug NLN 5200 M-260). The resulting driving magnetic field is
$H(t) = H_0 + \Delta H \sin{2\pi t f_D}$,
with $H_0$ as the
static and $\Delta H$ as the oscillating part of the magnetic field;
$f_D$ is the driving frequency.

At constant driving frequency $f_D$ and constant $\Delta H$ the
static part $H_0$ is increased in constant steps. After a relaxation
time of at least 5 s the minimum and maximum height is determined
for each value of $H_0$. The response period $T$ of the surface is
analysed by means of a correlation function.

\section{Experimental Results}

We modulate a subcritical bifurcation by means of an oscillating
magnetic field. The oscillation changes the character of this
bifurcation. The subsequent oscillations of the surface can be
harmonic, subharmonic or irregular depending on the magnitude of the
static and the amplitude and frequency of the oscillating magnetic
field. We condense the richness of the scenario into three
phase-diagrams,
using three representative driving frequencies.
The most complex behavior is observed at a driving frequency of 13 Hz.
The bifurcation diagrams simplify for smaller (larger) frequencies,
for which we have taken 2.5 Hz (23.5 Hz) as a representative.
These driving frequencies must be compared with the characteristic
time of the system, which is given by the decay time of the peak once
the field is turned off, and has been experimentally determined to be
about 40 ms.

\subsection{Modulating the subcritical bifurcation}

Fig.~\ref{height_exp} shows the height of the surface depending on
the static field $H_0$ without modulation (a) and with modulation
$\Delta H = 0.17 H_c$ and $f_D = 13$ Hz (b). The unit of the magnetic
field is the critical field $H_c = 6.2 \cdot 10^3$ Am$^{-1}$ of the
normal field
instability in the static case. The squares mark the increase
of the static field and the crosses mark the adjacent decrease.
Fig.~\ref{height_exp} (a) shows the hysteresis without field
modulation. Starting with a flat surface and increasing the magnetic
field leads to the onset of the normal field instability at
$H_0 = H_c$, producing a peak of 2.1 mm height. Decreasing the field
destroys the peak at the saddle node field $H_s = 0.94 H_c$.

In Fig.~\ref{height_exp} (b) the influence of the field modulation
$\Delta H = 0.17 H_c$ is presented. In order to characterize the
different states of the surface, including subharmonic behavior,
we need to observe the maximum height $h_{max}$ and the minimum
height $h_{min}$ of the surface during a sufficiently large time
period. Additionally the response period $T$ in units of the driving
period $T_D$ is shown. If it is not possible to detect a periodic
motion for a certain state, this state is marked with $T = 0$ in the
diagram.
We observe three qualitatively different oscillating states:
The first is the flat surface, where $h_{min}=0$ and $h_{max}=0$.
This state is labeled (00). It is measured for $H_0 < 0.92 H_c$.
For $H_0 > 1.08 H_c$ the peak oscillates around its equilibrium
height, that is $h_{min}>0$ and $h_{max}>0$. Therefore it is labeled
(++).
Between the flat surface and the oscillating peak lies a regime
($0.92 H_c < H_0 < 1.08 H_c$) where the peak periodically
arises and collapses to zero height. We label this behavior (0+).
In this regime the dynamics of the surface is determined by the ratio
of the driving period $T_D$ to the characteristic times a peak needs
to arise and collapse. This is the regime of interest where we
observe complex temporal behavior, involving subharmonics as
described below.
At $H_0 = 0.92 H_c$ the oscillating surface with response-period-1
develops softly from the flat surface. Deviations of $h_{max}$ between
the increasing and decreasing field are due to camera fluctuations.
The period-1-states bifurcate at $H_0 = 0.98 H_c$ into the
period-2-states in a supercritical way. The transition
from the period-2-states to the period-3-states at $H_0 = 1.015 H_c$
shows a small hysteresis in the maximum height $h_{max}$.
The coexisting period-2-attractor and
period-3-attractor cause an irregular motion. The peak then seems to
change its mode in an intermittent way. Further increase of $H_0$
produces the odd-numbered subharmonic cascade 3-5-7-9-11. The single
subharmonic states are separated by intermittent states. Further
increase of $H_0$ leads to the (++)-state of the oscillating peak.
We cannot exclude that the value of $h_{max}$ is finite for $H_0<0.92 H_c$.
In this range the $h_{max}$ is below our experimental
resolution. We do indeed expect a finite value of $h_{max}$ for any value
of the driving field due to the spatial inhomogeneity of the magnetic
field at the surface of the fluid.

The transition from the 1T to the 2T-mode is somewhat reminiscent of the
observation reported for a long channel in Ref.~3. In that case, as
indicated by their figure 1, the 2T- oscillation occurred in the form of
a standing wave very reminiscent of the Faraday instability, i.e. a
symmetric oscillation around the value $h$=0 was observed. In contrast,
the 2T-mode of the single peak in our experiment oscillates around a
finite positive value, and it appears via a period doubling bifurcation.

\subsection{Dynamical behavior}

An example of subharmonic behavior is demonstrated in
Fig.~\ref{period9}, where we observe a response-period $T = 9 T_D$.
The ordinate corresponds to the actual height of the peak, while
the abscissa indicates the time in units of the driving period. The
time interval $1T_D$ consists of 7 data points, which are linearly
interpolated. Deviations in the periodicity of the measure are due to
camera fluctuations.

\subsection{Phase-diagrams $\Delta H (H_0)$}

Figs.~\ref{H_dH_exp_1} - \ref{H_dH_exp_3} describe the
surface behavior by means of phase diagrams $\Delta H (H_0)$ at
three different driving frequencies. For each value of $\Delta H$
the static part $H_0$ is increased starting at such values of $H_0$
for which the surface is still flat. The phase-diagrams are
separated into the three domains (00), (0+) and (++).
The measured states of the flat (00)-domain are not marked with an
own color in the diagrams. But all states of the (0+)-domain are
shown in the diagrams by colors. In the (++)-domain only the
onset-state for each value of $\Delta H$ is shown.
The transition from (00) or (0+) to (++)-modes is indicated by a dotted
black line.
Each figure consists of two parts: (a) shows the color-coded
oscillatory mode and (b) shows the color-coded maximum height
$h_{max}$ of the surface. If it is not possible to identify a
certain periodicity the number is replaced by the character 'I',
because the peak then seems to change its subharmonic mode in an
intermittent way.

\subsubsection{Low frequency}

The results for low driving frequency $f_D = 2.5$ Hz are shown in
Figs.~\ref{H_dH_exp_1} (a) and (b). In the $\Delta H$-range from
0 to 0.023 $H_c$ we observe only two different states: the flat
surface (00) and the oscillating peak (++).
The threshold is decreased by modulation.
For $\Delta H > 0.023 H_c$ the (0+)-domain is observed, but the
response of the surface is always harmonic as can be seen in
Fig.~\ref{H_dH_exp_1} (a).
The existence of three qualitatively different domains is due to the
hysteresis of the static bifurcation. It can be understood in the
quasi-static limit of $f_D \to 0$.
Then we would expect the (00)-state if the total magnetic field
grew up from zero but stayed always below $H_c$, (0+)-modes if the
driving field were below the saddle node field $H_s$ for a certain
time of the period $T_D$, and (++)-modes if the driving field were
always above the saddle node field $H_s$. These two limit lines are
drawn in Fig.~\ref{H_dH_exp_1} (a): $\Delta H (H_0) = H_c - H_0$
(dashed line), $\Delta H (H_0) = H_0 - H_s$ (dashed-dotted line).
Fig.~\ref{H_dH_exp_1} (b) shows the maximum heights $h_{max}$ of the
peak. The higher $H_0$ the higher $h_{max}$, which is in accordance
with the idea of the quasi-static limit.

\subsubsection{Medium frequency}

Fig.~\ref{H_dH_exp_2} shows the results for $f_D = 13$ Hz.
For $\Delta H < 0.08 H_c$ we measure only two different states,
similar to the behavior at a frequency of 2.5 Hz:
the flat surface (00) and the oscillating peak (++). The
threshold is slightly increased by modulation in contrast to the
measurement at 2.5 Hz. The $\Delta H$-range of the direct
transition between the two domains is also increased.
For $\Delta H > 0.08 H_c$ there exists the (0+)-domain again, but now
this domain has a fine-structure of different oscillatory modes as
can be seen in Fig.~\ref{H_dH_exp_2} (a). For low values of $H_0$
the surface response is harmonic. Increasing $H_0$ leads to a
period-2-state. For sufficiently high oscillating fields,
$\Delta H > 0.2 H_c$, and for high values of $H_0$ this state
converts directly into the (++)-domain. But in the $\Delta H$-range
from 0.09 to 0.2 $H_c$ we observe a 'balloon' of subharmonic and
intermittent oscillations. For lower values of $H_0$ the odd-numbered
subharmonic cascade of periods 3,5,7 and 9 $T_D$ can be observed
inside this balloon. These different states seem to be separated
by intermittent states caused by coexisting attractors
\cite{Lauterborn91,Lauterborn93} of the pure states.
While a period-1-state could
softly bifurcate into a period-2-state, there exists no smooth
transition from $T=2$ to $T=3$ and we observe irregularly
change of the dynamics between these attractors.
The diagram shows that the basins of attraction
are getting smaller for higher modes. For higher values of $H_0$ we
can only observe intermittent states and the even-numbered
subharmonic modes 4,6,8,10,12 and 16 within the balloon. 
A regular band-structure, like the odd-numbered cascade, of the
even-numbered modes can not be found in the phase-diagram.
The maximum height $h_{max}$ of the peaks is shown in
Fig.~\ref{H_dH_exp_2} (b). The transitions  from period-1 to
period-2, from period-2 to the balloon, and from the balloon to
period-2 are indicated by small black bars. We observe, that some
changes of the oscillating modes are connected with changes of
$h_{max}$. At the transition point from period-1 to period-2
$h_{max}$ is increased. The period-2 to period-3 transition also
shows an increase of $h_{max}$.
But in the small intermittent band between the period-2-states and
the period-3-bands the height is decreased.
There exist local maxima of $h_{max}$ inside the period-1-band and the
period-2-band, indicating the areas of strong resonance of the
oscillator.

\subsubsection{High frequency}

In Fig.~\ref{H_dH_exp_3} the results for high driving frequency
$f_D = 23.5$ Hz are presented. The $\Delta H$-range from 0 to
0.14 $H_c$, where we find only the flat surface and the domain with
$h_{max}>0$, is larger than in the case of 13 Hz or 2.5 Hz. This can
be understood, because in the case of $f_D \to \infty$ the fluid
should not be influenced by the oscillating part. Therefore the
(0+)-domain appears at higher values of $\Delta H$ for the high
frequency $f_D = 23.5$ Hz.
The threshold for the transition between the flat surface (00) and
the (++)-domain is increased.
For $\Delta H > 0.14 H_c$ we observe only even-numbered
oscillating modes 2,4,6,8 and 10, the period-1-state and the
intermittent state as shown in Fig.~\ref{H_dH_exp_3} (a). For low
values of $H_0$ there exists only the period-2-state, except for two
data points. The period-2-state changes into the period-4-state
by a period-doubling-bifurcation in the $\Delta H$-range from 0.23
to 0.29 $H_c$.
For $\Delta H = 0.29 H_c$ the period-4-state is converted into a
period-8-state if $H_0$ is increased.
Inside the period-2 domain there exists a balloon again consisting of
states with period 4,6 and 8 $T_D$ as well as intermittent states.
The maximal height $h_{max}$ is shown in Fig.~\ref{H_dH_exp_3} (b).
The transition from period-2 to period-4 or intermittency is
again indicated by small black bars.
Except for the area of the diagram described by high values of
$\Delta H$ and low values of $H_0$ the height of the peaks is
decreased if $H_0$ is increased, in contrast to the cases of lower
driving frequencies $f_D$.

Even- and odd-numbered subharmonic responses are known from
simulations of driven nonlinear oscillators
\cite{Lauterborn91,Lauterborn93}.
Their range of existence is governed by the ratio of the driving period
$T_D$ to the natural period of the
oscillator, which is in our case the time a peak needs to arise and
collapse.
In order to achieve a deeper understanding of the existence of
the odd-numbered subharmonic cascade we derive a theoretical minimal
model which reproduces the most striking observations of the
bifurcation scenario.

\section{Minimal Model}

We present a minimal theoretical model which captures the essence of
the experimental findings, namely the existence of two oscillatory
domains, the subharmonic dynamics, the hysteresis, the threshold
shift and the static behavior
\begin{eqnarray*}
\ddot{h} + \beta \dot{h} & = & h - h^2 + \epsilon(t) \\
             \epsilon(t) & = & H_0 + \Delta H \sin{2\pi t/T_D}-H_c \\
\end{eqnarray*}
with a cutoff condition: $\dot{h}$ and $h$ are set to zero if $h$
reaches negative values.
$h$ represents the height of the peak and $\beta$ measures the
damping. We use a second derivative in the equation because the driving
period is not very different from the characteristic times that
a peak needs to arise and collapse.
The force $h - h^2$ is the minimal form which considers the
asymmetry of the system and which causes a hysteresis in connection
with the cutoff condition.
The additive coupling of the control parameter $\epsilon$ represents the
magnetic field inhomogeneity caused by the magnetization of the fluid.
Near the rim of the vessel, the field lines are not normal to the fluid
surface. Thus, even small subcritical fields will lead to deformations of
the fluid surface. Strictly speaking, the rising of the peak does not
stem from a bifurcation, and the additive coupling takes that into account.
The cutoff condition is in accordance with the observation of very
strong dissipative forces caused by the boundaries of the vessel when
the peak breaks down.

Figs.~\ref{height_theo} (a) and (b) show the results of the numerical
simulation for the parameters obtained from the fit explained below.
At fixed driving the static part $H_0$ is
increased from 0.7 to 1.2 $H_c$. For each value of $H_0$ the
dynamic is determined after relaxation and is marked in the
diagrams with a square. Crosses mark the adjacent decrease of
$H_0$.
Fig.~\ref{height_theo} (a) demonstrates the hysteresis of the
static case without modulation. At $H_0 = H_c$ the solution $h=0$
becomes unstable and the height $h$ jumps to the fixpoint
$h^\star (\epsilon=0) = 1$. Further increase of $H_0$ shifts
the fixpoint to
$h^\star (\epsilon) = \sqrt{\epsilon + \frac{1}{4}} + \frac{1}{2}$.
Decreasing $H_0$ leads to the saddle node
$H_s = H_c - \frac{1}{4}$ where the finite solution becomes unstable.
This dynamical behavior corresponds to the measured dynamic of
Fig.~\ref{height_exp} (a).
The influence of modulation is shown in Fig.~\ref{height_theo} (b).
Starting again at the (00)-state at low values of $H_0$, the solution
$h=0$ becomes unstable in a supercritical way at $H_0 < H_c$. The
response period of the (0+)-state is $T = T_D$. By increasing $H_0$
this mode transforms supercritically into a period-2-state. Further
increase of $H_0$ leads to higher subharmonic modes and intermittent
states, which are marked in the diagram again with $T = 0$.
For $H > 1.1 H_c$ the (++)-state becomes stable. A similar dynamic is
observed in the experiment in Fig.~\ref{height_exp} (b). A decrease
of $H_0$ shows no hysteresis except for some transitions in the
subharmonic and intermittent regime.

In the model the exact transition from (00) to (0+)-domains is
always determined by the line $\Delta H (H_0) = H_c - H_0$, because
then the resulting force $h-h^2+\epsilon(t)$ is positive for a certain
time. According to an experimental resolution limit of the
height we introduce a threshold height $h_{res}$ for the presentation
of the simulations in the phase-diagram $\Delta H (H_0)$ of
Fig.~\ref{H_dH_theo}.
The model-parameters $\beta$, $T_D$, $H_c$ and $h_{res}$ are used as
fit-parameters to fit the numerical transition-line (00) $\to$
(0+) to the transition-line from the measurement at 13 Hz presented
in Fig.~\ref{H_dH_exp_2} \cite{NumRec}. The initial parameters of
each fit are chosen randomly from a certain interval. The best fits
show good agreement with the measured transition-line (00) $\to$
(0+), but the inner structure of the (0+)-domain does not match with
the measurements. Therefore in Fig.~\ref{H_dH_theo} there is presented
the best fit, which shows an inner structure of the (0+)-domain
similar to the experimental data:
$\beta = 0.07825422$,
$T_D = 2.87910700$, $H_c = 7.48344946$ and $h_{res} = 0.4278$.
The representation of the data is the same as in
Figs.~\ref{H_dH_exp_1} (a) -- ~\ref{H_dH_exp_3} (a), except for the
label 'I', which marks periodic states with $T > 10 T_D$ and states
where no periodicity can be detected.
The solid black line in Fig.~\ref{H_dH_theo} corresponds to the
dotted black lines in Figs.~\ref{H_dH_exp_1} -- ~\ref{H_dH_exp_3},
indicating the transition from (00) or (0+) to (++)-modes.
In the $\Delta H$-range from 0 to 0.1 $H_c$ the (00)-domain
transforms directly into the (0+)-domain.
For $0.050 H_c < \Delta H < 0.226 H_c$ we observe that (++)-states
can retransform into (0+)-states in contradiction to the experimental
data shown before. Although this feature does not occur at all
combinations of the model-parameters, we suppose that the
retransformation could be suppressed by higher non-linear terms.
As in the experimental case referring to Fig.~\ref{H_dH_exp_2} (a) we
detect mainly the subharmonic cascade of response periods 1,2,3,5,7,9
inside the (0+)-domain for lower values of $H_0$.
The period-2-regime includes a balloon of higher modes and
intermittent states. The basins of attraction become smaller
for higher modes in accordance with the experimental findings.

\section{Summary and Conclusion}

We have studied the nonlinear surface oscillations of a magnetic fluid
under the influence of a time dependent magnetic field in the
neighborhood of a subcritical bifurcation. The nonlinear response
involves subharmonic regimes with periods up to 11 $T_D$, which are
separated by regimes of irregular oscillations.
At low frequencies the surface follows the frequency of the driving.
At high frequencies only even numbered subharmonics can be observed,
while at medium frequencies a regime arises where odd numbered
subharmonics are dominant. These features are captured by a minimal
model of the nonlinear oscillator.

We have used a harmonic driving
$H(t) = H_0 + \Delta H \sin{2\pi t f_D}$. It must be kept in
mind that a time-periodic driving with different ratios of harmonics,
i.~e. square waves, will lead to quantitative differences in the
bifurcation scenario. We have not performed a systematic study of those
influences.

It seems to be interesting to study the interaction of many spatially
coupled oscillating peaks. In the regime where subharmonic responses
are favored one would expect spatial domains with different phases of
the oscillation. Thus experimental investigations in larger vessels
are currently under way.

\section{Acknowledgment}

We would like to thank S.~Linz and U.~Parlitz for stimulating
discussions.
The experiments are supported by the 'Deutsche
Forschungsgemeinschaft' through Re588/10.

\begin{figure}[p]
\caption {\label{setup} Experimental setup.}
\end{figure}

\begin{figure}[p]
\caption {\label{minmaxFoto}
Two snapshots of the ferrofluid peak in the vessel. The first one is
taken at the phase of the oscillation where the amplitude reaches its
minimum value, the second one taken at the maximum amplitude.
}
\end{figure}

\begin{figure}[p]
\caption {\label{height_exp} a) $\Delta H = 0$: Measurement of
the height of the peak as a function of the static field $H_0$.
b) $\Delta H = 0.17 H_c$: Measurement of the response-period
$T$ in units of the driving period $T_D$, and measurement of the
maximum and minimum height.
The driving period $T_D$ is $76.16$ ms. The upper part
of each diagram corresponds to the minimum and maximum height of
the peak in mm.
Squares (crosses) correspond to the increasing (decreasing) field.}
\end{figure}

\begin{figure}[p]
\caption {\label{period9} shows the height of the peak during
a time of 18 periods of excitation at $f_D = 13$ Hz.
The solid line is obtained by a harmonic interpolation using the
frequencies up to $3.5/T_D$, whose amplitude is determined by means of a
discrete Fourier transformation.
The time interval $1T_D$ consists of 7 phase-locked sampled data points.
Thus small deviations in the periodicity  of the measure are
presumably due to camera fluctuations.
(For this measure we used a mixture of EMG 901 and EMG 909 in a
ratio of 4 to 1, in contrast to the ratio of 7 to 3, in order to
obtain a larger height of the peak.)
}
\end{figure}

\begin{figure}[p]
\caption{ \label{H_dH_exp_1}
Characterization of the measured surface dynamics depending on
the increasing static field $H_0$ and the oscillating field
$\Delta H$ at the low driving frequency $f_D = 2.5$ Hz;
$H_c = 6.6 \cdot 10^3$ Am$^{-1}$.
The transition from (00) or (0+) to (++)-modes is indicated by a
dotted line.
a) Only harmonic response is observed (green).
Stationary limit: $\Delta H (H_0) = H_c - H_0$ (dashed line),
$\Delta H (H_0) = H_0 - H_s$ (dashed-dotted line).
b) Coding of the maximum height.}
\end{figure}

\begin{figure}[p]
\caption{ \label{H_dH_exp_2}
Characterization of the measured surface dynamics depending on
the increasing static field $H_0$ and the oscillating field
$\Delta H$ at the driving frequency $f_D = 13$ Hz;
$H_c = 6.2 \cdot 10^3$ Am$^{-1}$.
The transition from (00) or (0+) to (++)-modes is indicated by a dotted
black line.
a) Color-coding of the response period of the surface.
The subharmonic cascade 1,2,3,5,7 and 9 $T_D$ is observed.
b) Coding of the maximum height.}
\end{figure}

\begin{figure}[p]
\caption{ \label{H_dH_exp_3}
Characterization of the measured surface dynamics depending on
the increasing static field $H_0$ and the oscillating field
$\Delta H$ at the driving frequency $f_D = 23.5$ Hz;
$H_c = 6.7 \cdot 10^3$ Am$^{-1}$.
The transition from (00) or (0+) to (++)-modes is indicated by a dotted
black line.
a) Color-coding of the response period of the surface.
Only even-numbered subharmonic response is observed.
b) Coding of the maximum height.}
\end{figure}

\begin{figure}[p]
\caption {\label{height_theo}
Numerical simulation for the parameters obtained from the fit:
a) Height of the peak depending on the static field $H_0$ without
modulation.
b) Minimum and maximum height of the peak and response period $T$
depending on the static field $H_0$ at $\Delta H = 0.24 H_c$.
Squares (crosses) correspond to the increasing (decreasing) field. }
\end{figure}

\begin{figure}[p]
\caption {\label{H_dH_theo}
Characterization of the surface dynamics obtained by numerical
simulation for the parameters resulting from the fit.
For each value of $\Delta H$ the static part
$H_0$ is increased from 0.8 to 1.2 $H_c$. The response periods are
shown by colors. The solid black line indicates the transition from
(00) or (0+) to (++)-modes. States lying on this line are already
(++)-states.
}
\end{figure}

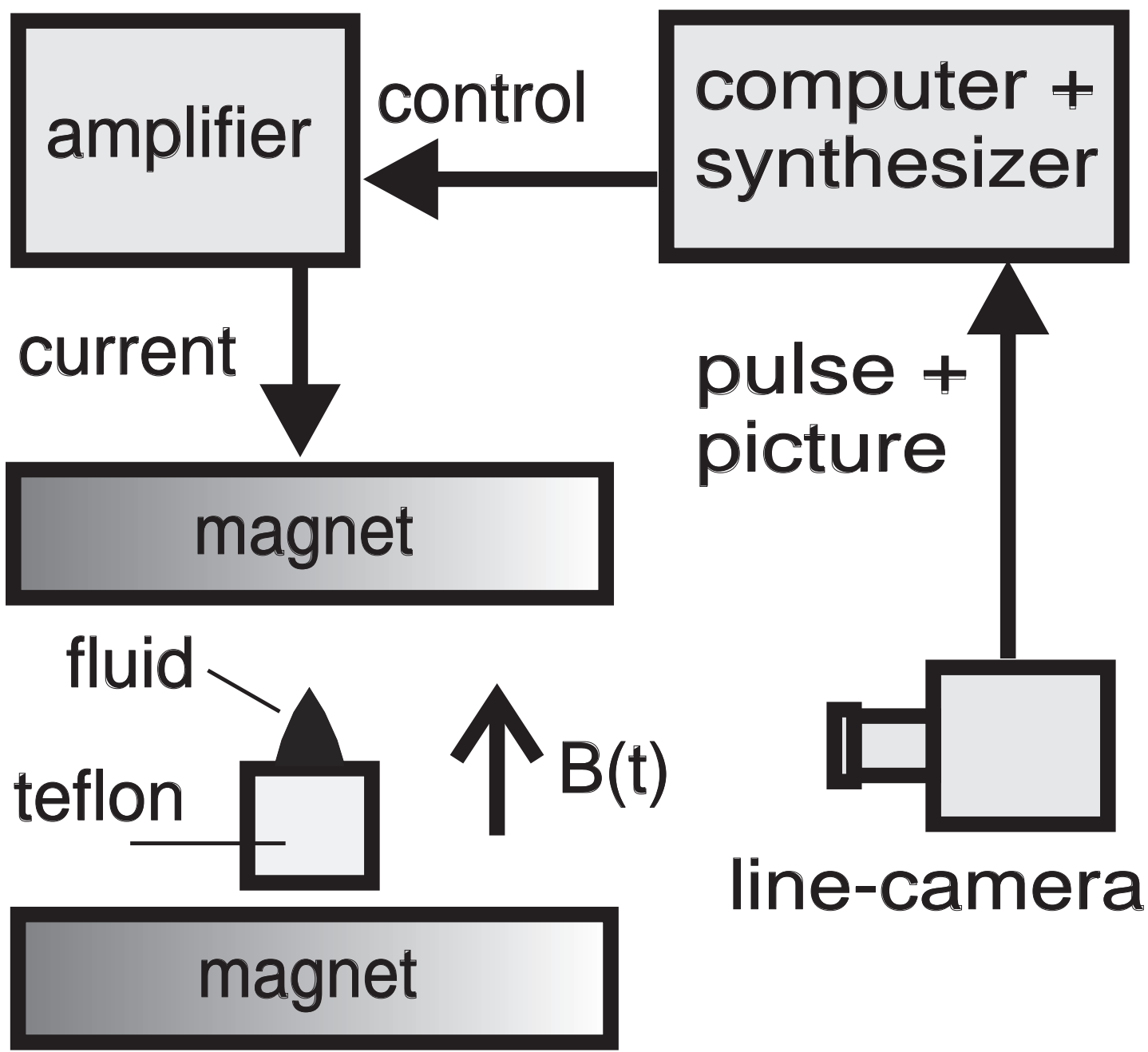
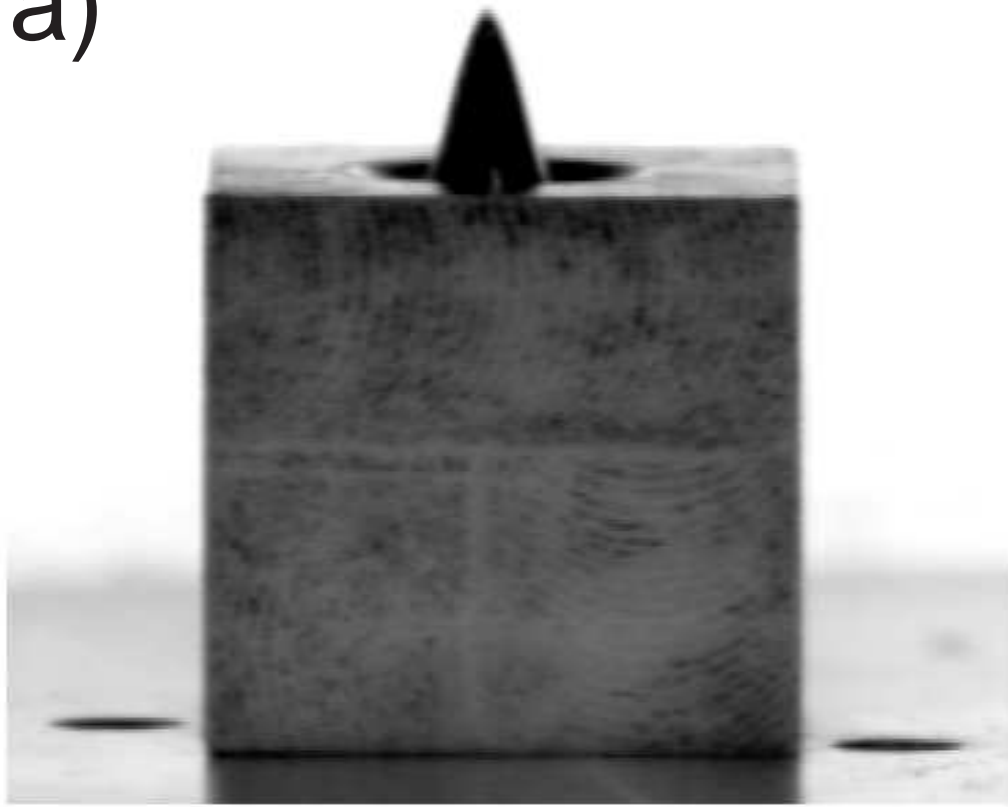
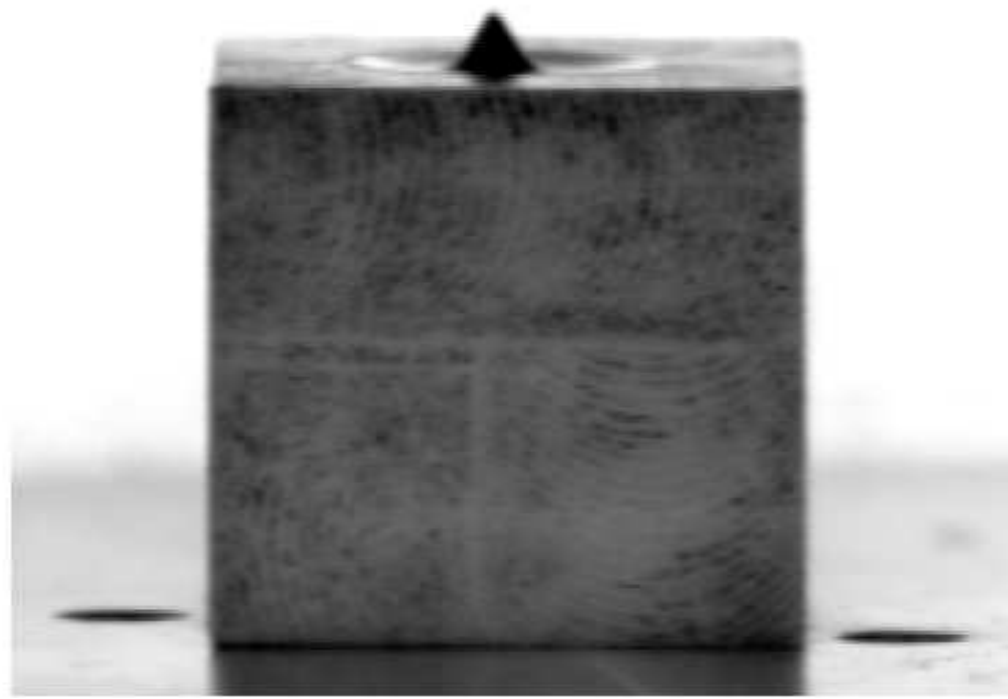
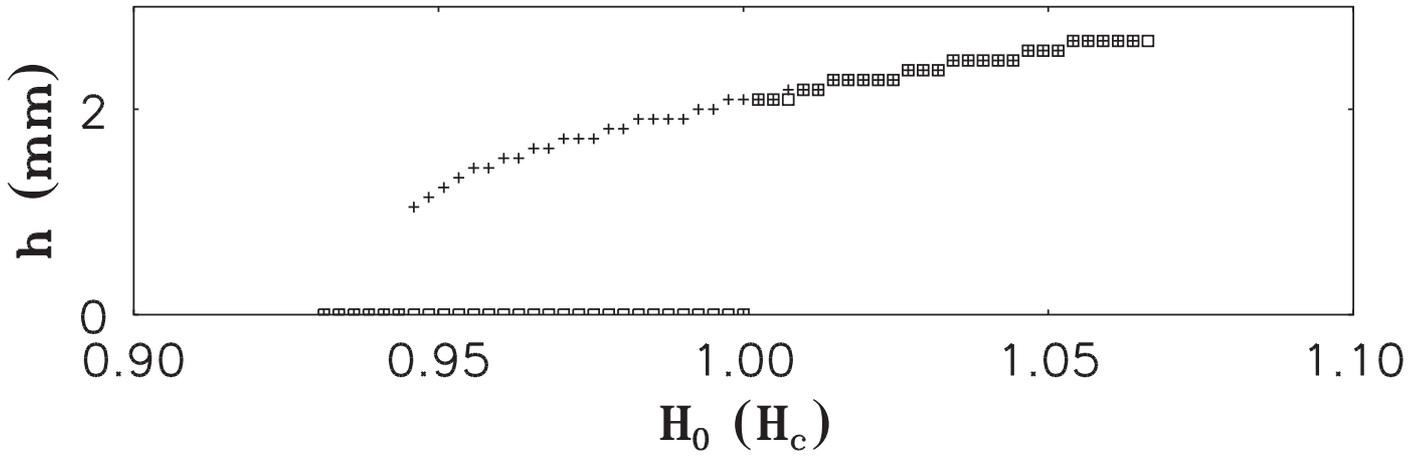
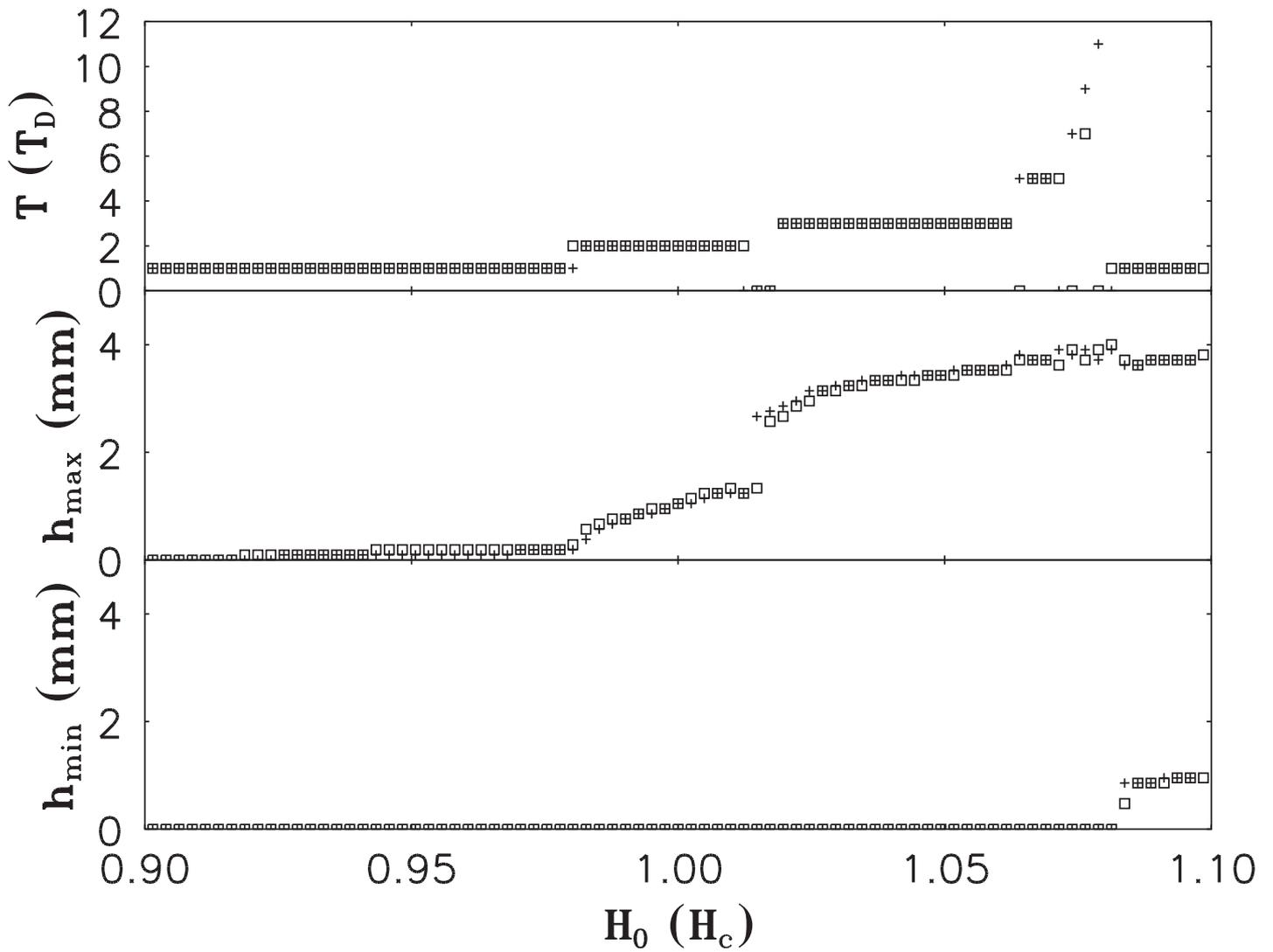
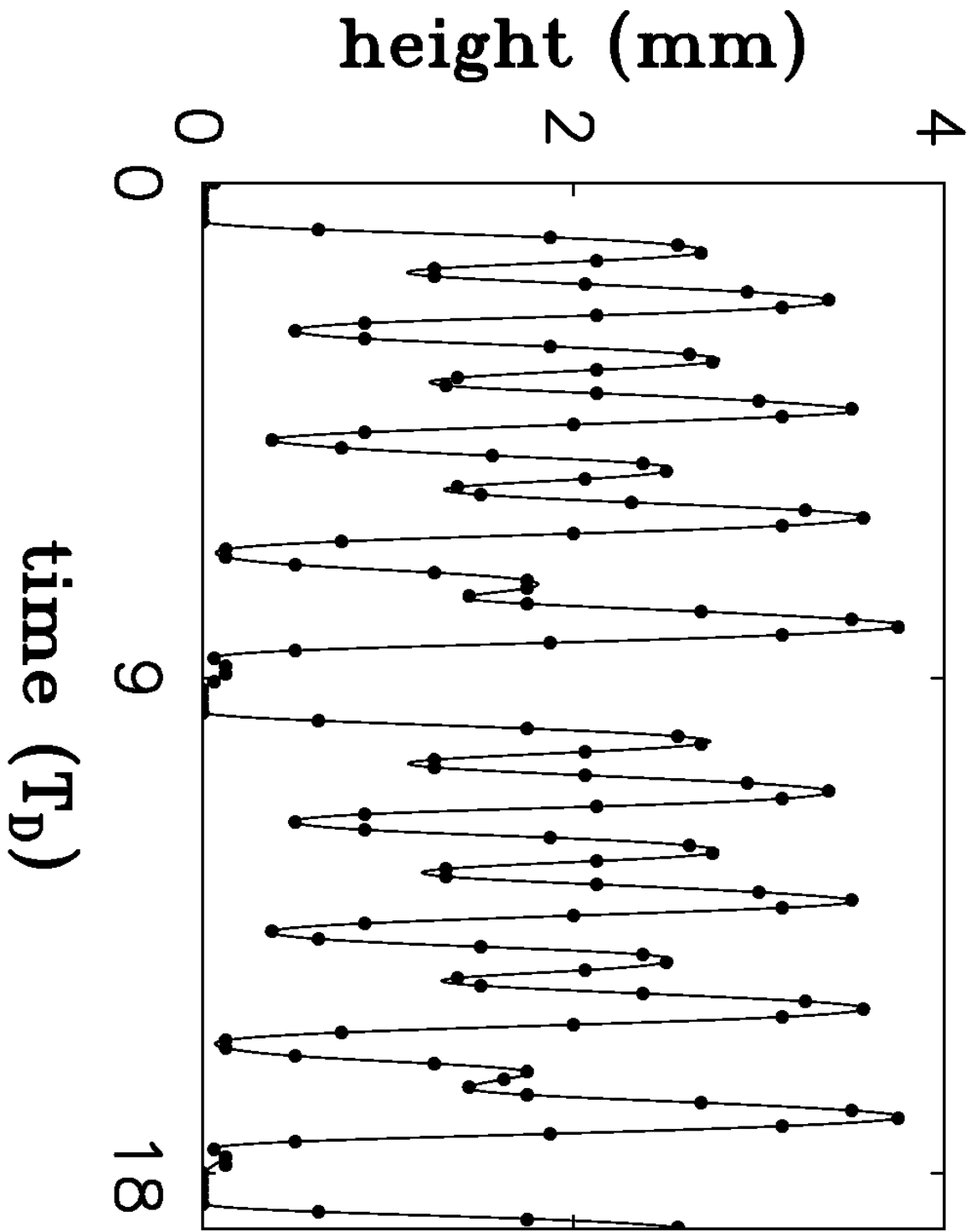
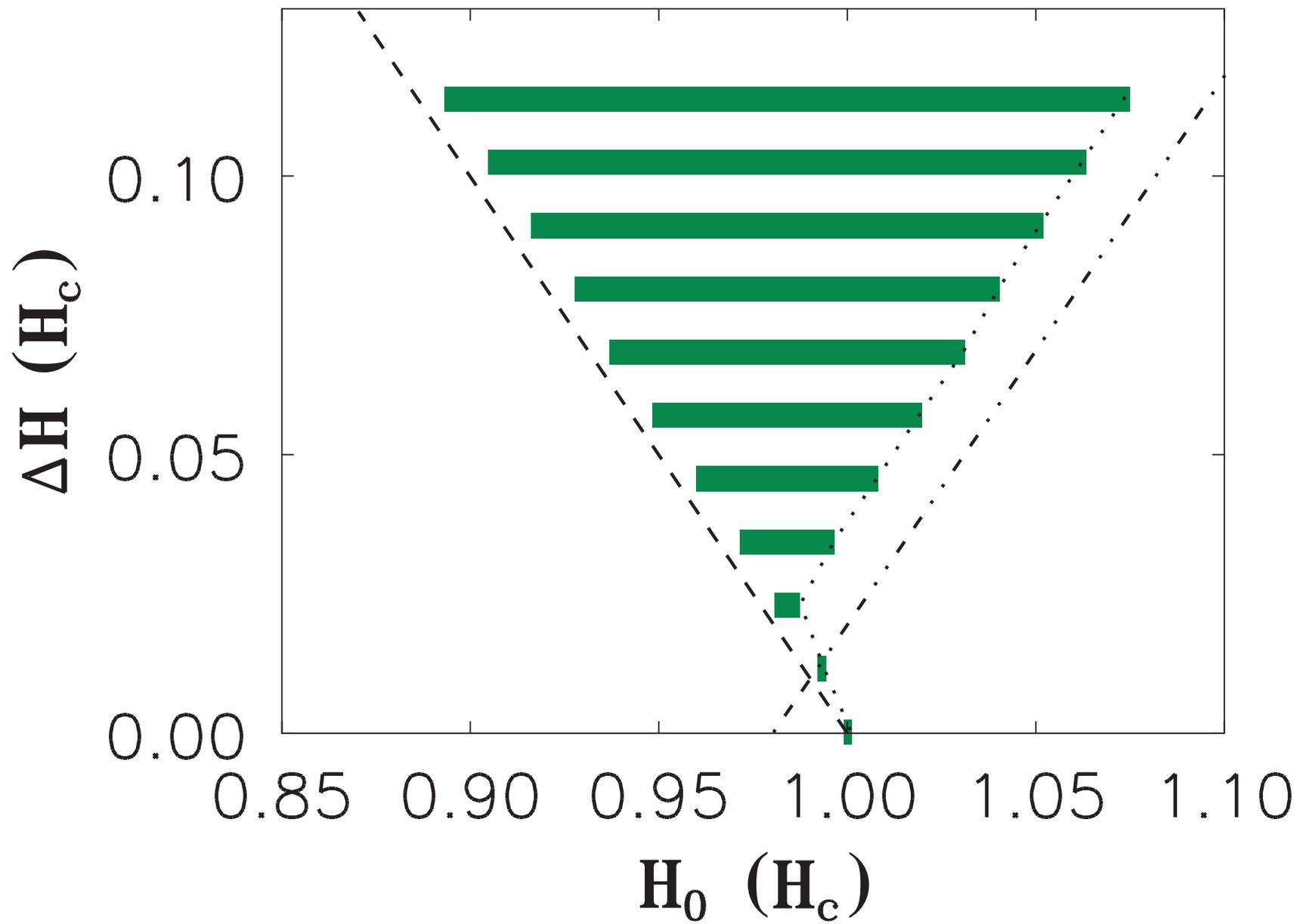
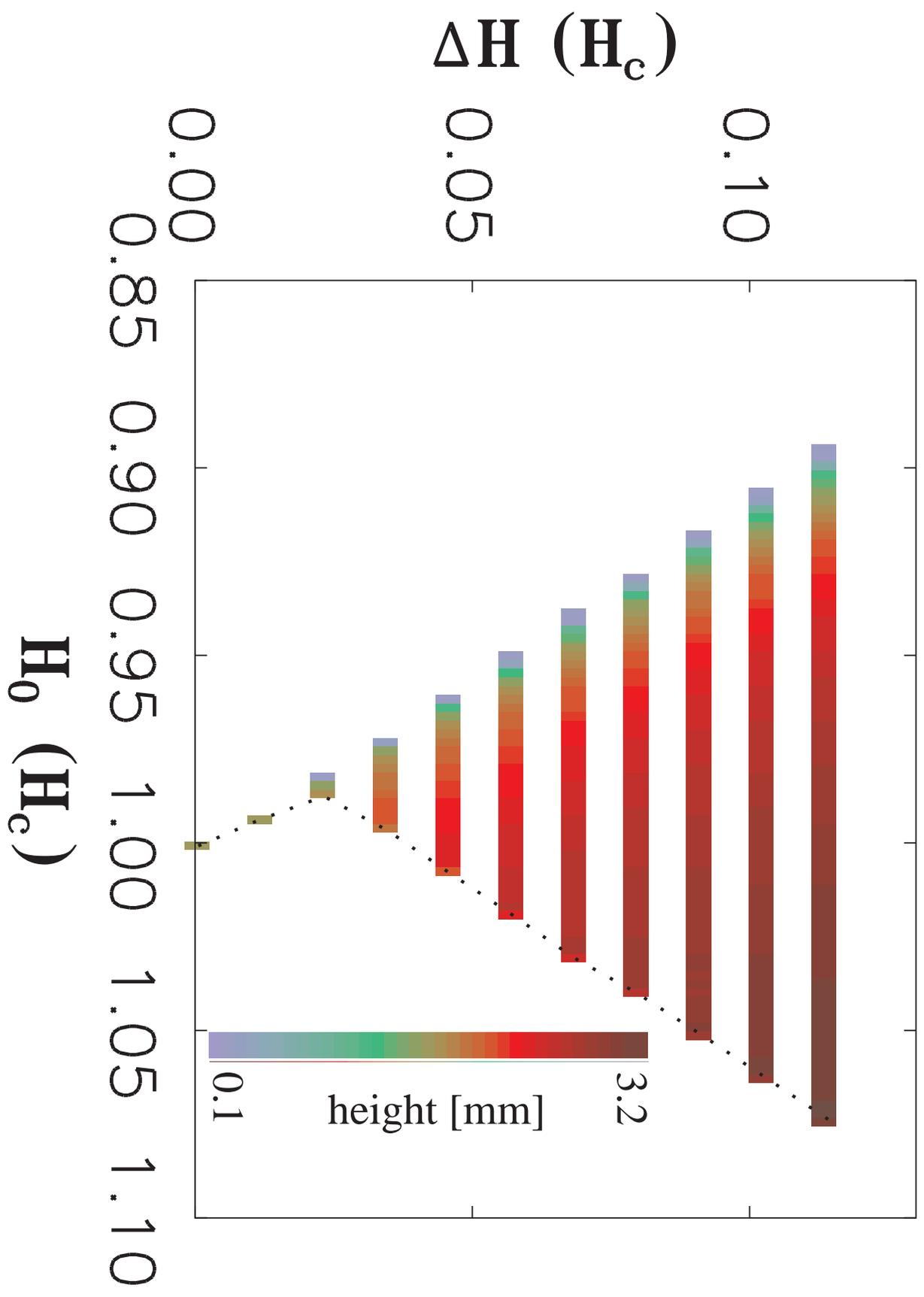
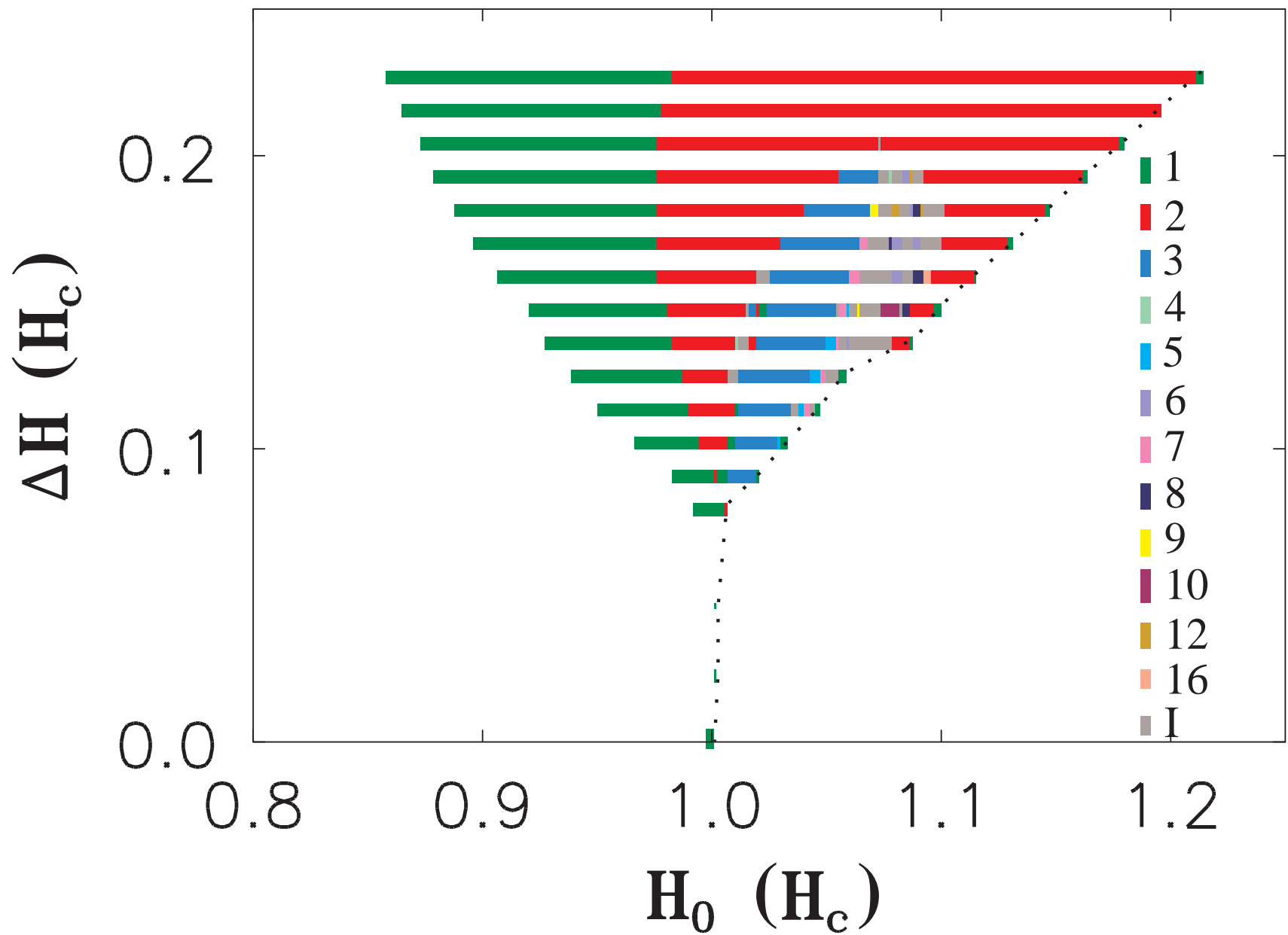
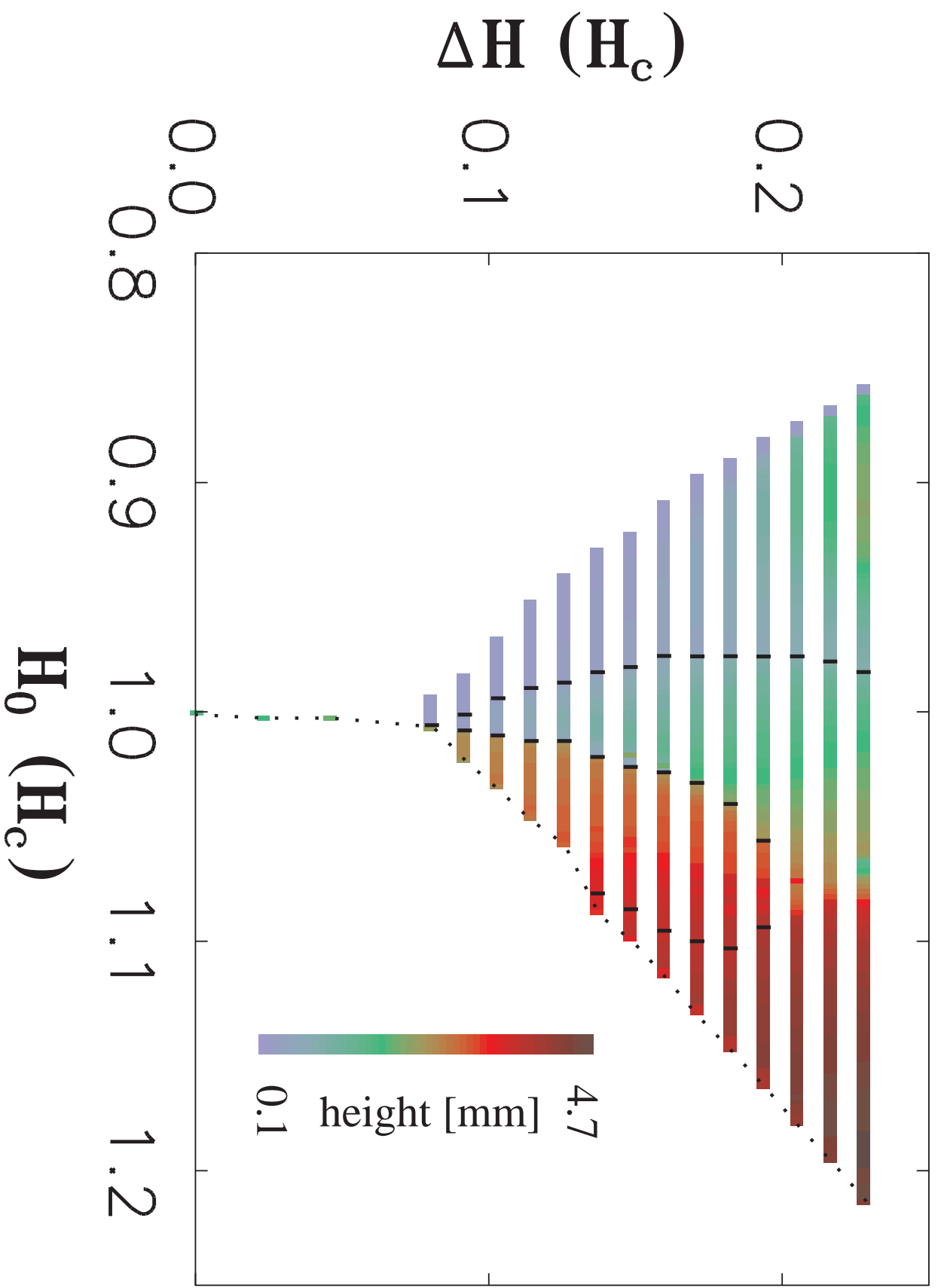
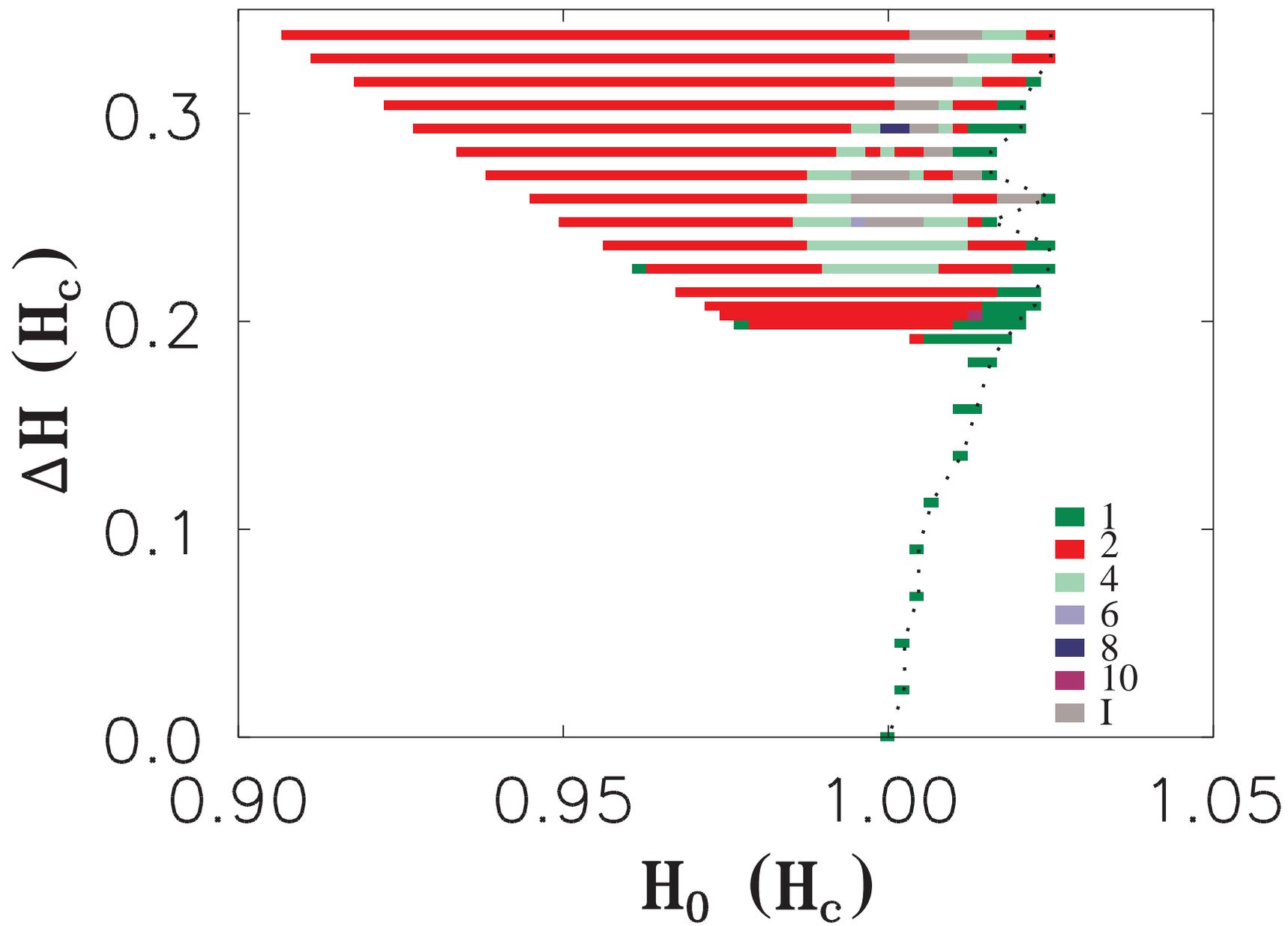
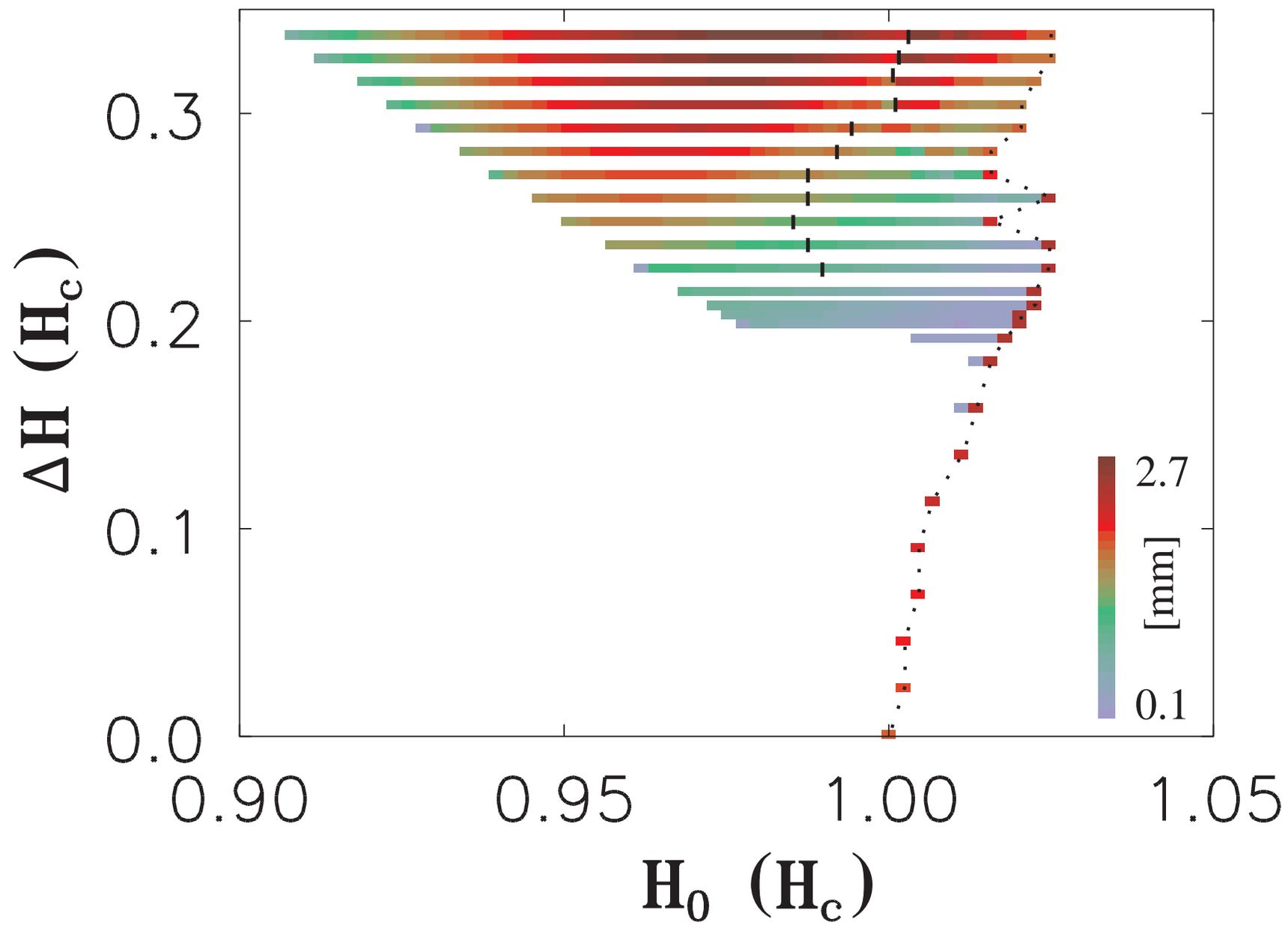
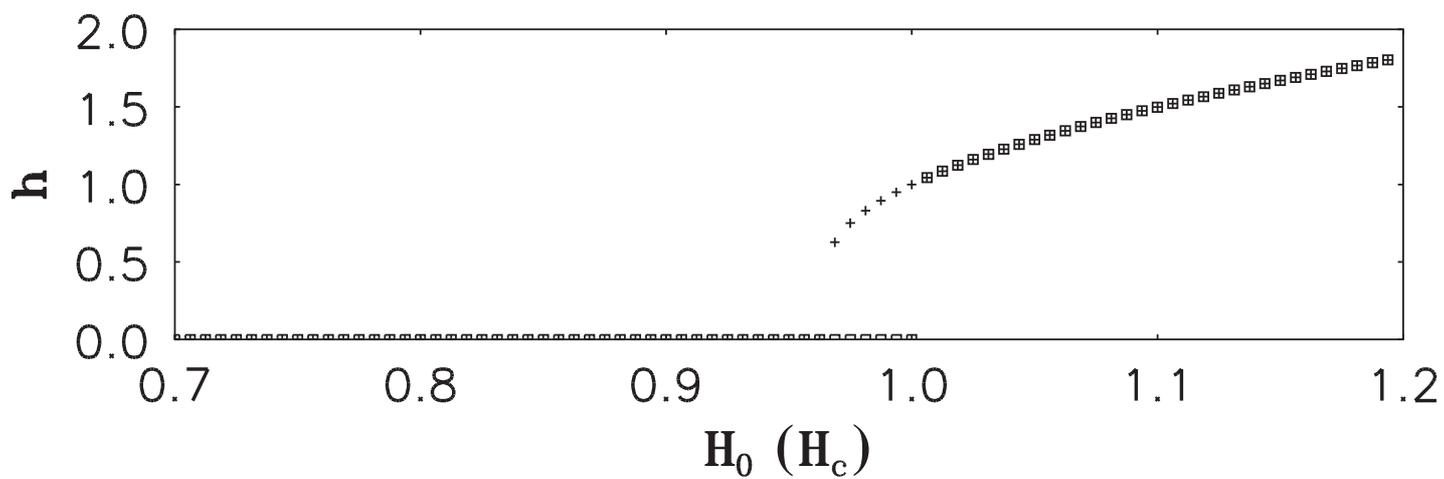
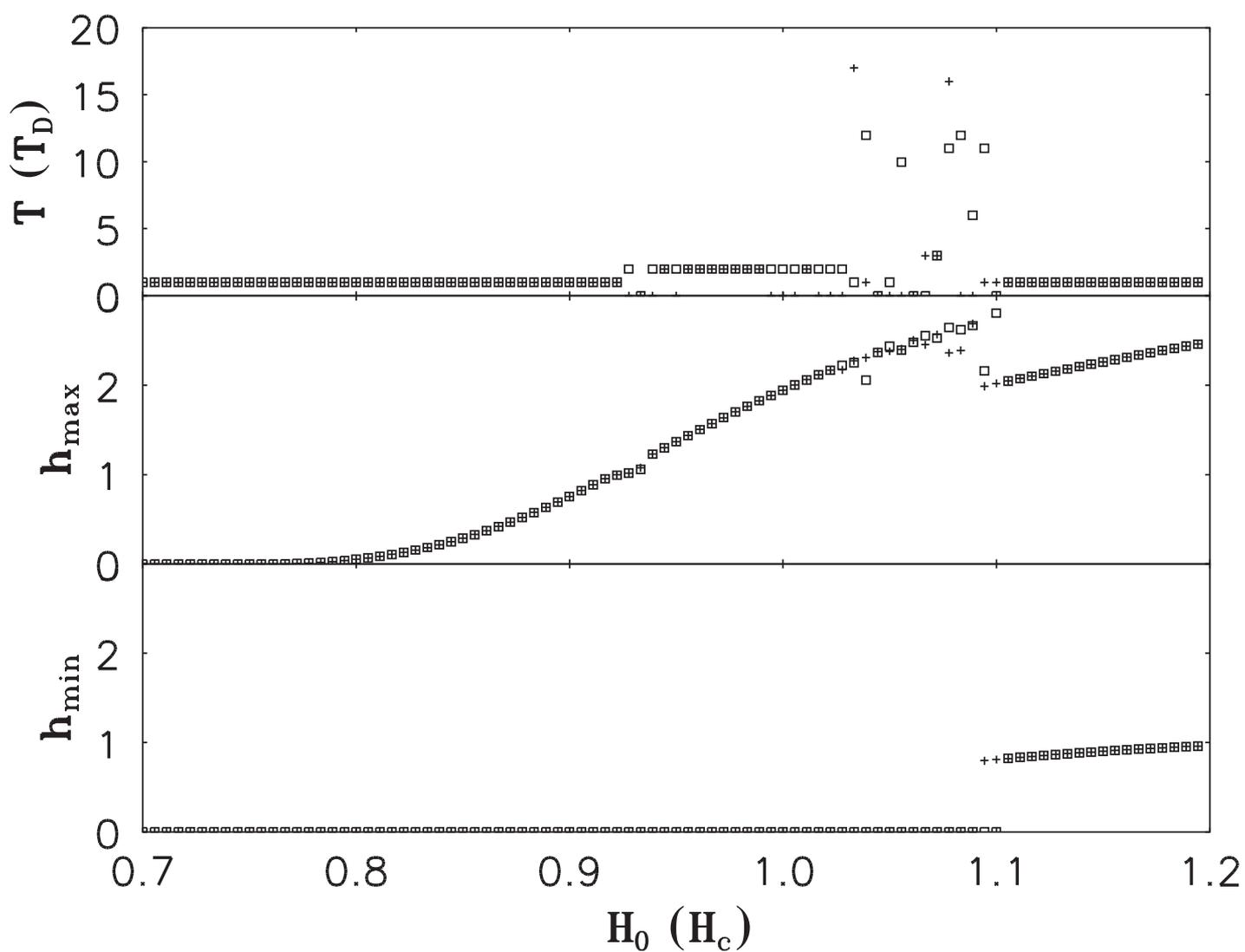
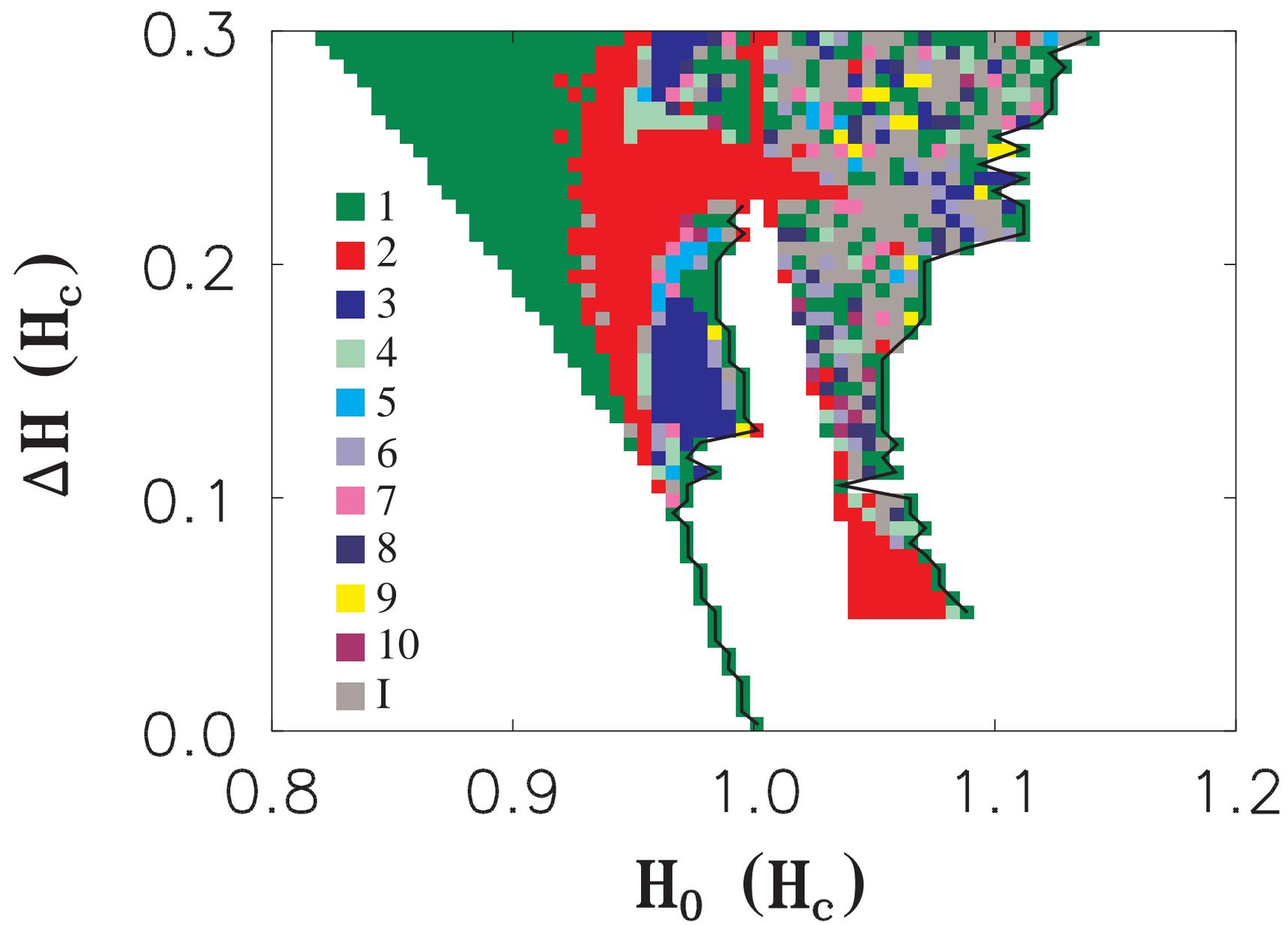
\end{document}